\documentclass[aps,twocolumn,showpacs,superscriptaddress]{revtex4}
\usepackage{amsmath}
\usepackage{amssymb}
\usepackage{epsfig}
\usepackage{bm}

\newcommand{\half} {\frac{1}{2}}
\newcommand{\e} { {\rm e} }

\newcommand {\ys} {\left|y_s\right|}
\newcommand {\scharge} {\tilde{\sigma}}

\begin{document}


\title {Adsorption and Depletion of Polyelectrolytes from
    Charged Surfaces\\  }

\author {Adi Shafir}
\email{shafira@post.tau.ac.il}
\affiliation{School of Physics and Astronomy   \\
    Raymond and Beverly Sackler Faculty of Exact Sciences \\
    Tel Aviv University, Ramat Aviv, Tel Aviv 69978, Israel\\}

\author{David Andelman}
\email{andelman@post.tau.ac.il}
\affiliation{School of Physics and Astronomy   \\
    Raymond and Beverly Sackler Faculty of Exact Sciences \\
    Tel Aviv University, Ramat Aviv, Tel Aviv 69978, Israel\\}
\author{Roland R. Netz}
\email{netz@theorie.physik.uni-muenchen.de}
\affiliation{Sektion Physik, Ludwig-Maximilians-Universit\"{a}t\\
        Theresienstr. 37, 80333 Munich, Germany\\}

\bigskip
\date{April 15, 2003}
\bigskip

\begin{abstract}
\setlength{\baselineskip} {14pt} \linespread{1.6}

Mean-field theory and  scaling arguments are presented to model polyelectrolyte
adsorption from semi-dilute solutions onto charged surfaces. Using
numerical solutions of the mean-field equations, we show that
adsorption exists only for highly charged polyelectrolytes in low
salt solutions. Simple scaling laws for the width of the adsorbed
layer and the amount of adsorbed polyelectrolyte are obtained.  In
other situations the polyelectrolyte chains will deplete from the
surface. For fixed surface potential conditions, the salt
concentration at the adsorption--depletion crossover scales as the
product of the charged fraction of the polyelectrolyte $f$ and the
surface potential, while for a fixed surface charge density,
$\sigma$, it scales as $\sigma^{2/3}f^{2/3}$, in agreement with single-chain results.\\

\end{abstract}

\maketitle
\newpage
\section{Introduction}
\label{Intro} \setlength{\baselineskip} {14pt} \linespread{1.6}

The phenomenon of adsorption of charged polymer chains (polyelectrolytes)  to
surfaces has generated a great
deal of interest due to its numerous industrial applications and
relevance to biological systems. The theoretical treatment
is not yet well established because of the multitude of length
scales involved, arising from different interactions:
electrostatic interactions between monomers and counter-ions,
excluded volume interactions and entropic considerations. Furthermore, when
salt is added to the solution, the interplay between polyelectrolytes (PEs) and salt
ions  as well as the ion entropy has to be taken into account.

The adsorption of PE chains
onto charged surfaces has been addressed theoretically in
several models in the past. They include among others:
solutions of linearized mean-field
equations
\cite{Varoqui,Varoquietal,Chatellier,Joanny,Manghi,Wiegel,Muthukumar},
numerical solutions of full mean-field equations
\cite{Itamar,Itamar1,Itamar2}, various scaling theories for
single-chain adsorption \cite{Borisov,Dobrynin},
and formulation of a phenomenological criterion describing
the adsorption--depletion transition
from charged surfaces \cite{NetzJoanny,review,review1}. Other approaches
employed multi-Stern layer models
\cite{vanderSchee,vandestig,Bohmer}, where a discrete lattice is
used and each lattice site can be occupied by either a monomer, a
solvent or a small ion. The electrostatic potential can then be
calculated self-consistently together with the concentrations of the
monomers and counterions.

In this article we re-examine the mean-field equations describing
the PE adsorption and their
numerical solutions, with
specific emphasis on the adsorption--depletion transition.
The present paper
can be regarded as an extension of Ref. \cite{Itamar}. It
agrees with the previously obtained low-salt adsorption regime
but proposes a different interpretation of the high-salt regime.
We find that the high-salt adsorption regime of Ref.
\cite{Itamar} is pre-empted by an adsorption--depletion transition, in analogy with
single-chain results.
The mean-field equations and their numerical solutions
are formulated in Sec. \ref{EQEQ}, some simple
scaling relationships in Sec. \ref{SCale}, and the
adsorption--depletion transition in Sec. \ref{Phaset}. A
general discussion and comparison with other models are presented
in Sec.~\ref{comparison}.

\section{The Mean Field Equations and Their Numerical Solution}
\label{EQEQ}
\linespread{1.6}

Consider an aqueous solution of infinitely long PEs, together
with their counterions and an added amount of salt. Throughout
this paper we assume that both the salt ions and counterions are
monovalent. Let ${\phi}({\bf r})=\sqrt{c({\bf r})}$ be the square root of $c({\bf r})$,
the
local monomer concentration, $a$ the monomer size and $f$ the
charge fraction on each PE chain. Also let $\phi_{\rm b}=\sqrt{c_{\rm b}}$ be the
square root of the bulk monomer concentration $c_{\rm b}$, and $\psi({\bf r})$ the
electrostatic potential. The mean-field free energy can be obtained
either from phenomenological  or  field theoretical approaches:
\nolinebreak[2]
\begin{equation}
 F= \int {\rm d}  {\bf {r}} \left(f_{\rm pol}+f_{\rm ion}+
        f_{\rm el}\right)
\label{F} 
\end{equation}
\begin{eqnarray}
  f_{\rm pol}= 
        k_{\rm B} T\left(\frac{a^2}{6}\left|\nabla \phi \right|^2+\half v
    \left(\phi^4-\phi_{\rm b}^4\right)-\mu \left(\phi^2-\phi_{\rm b}^2\right)
    \right)
\label{fpol}
\end{eqnarray}
\begin{eqnarray}
 f_{\rm ion} = k_{\rm B} T \sum_{i=+,-}\left[c^i\ln c^i-c^i-\mu^i\left
    (c^i-c_{\rm b}^i\right)\right]
\label{fion} \\
 f_{\rm el}=\left(c^+ - c^- +f\phi^2\right)e\psi-
    \frac{ \epsilon}{8\pi}\left|\nabla \psi\right|^2 ~.
\label{fel}
\end{eqnarray}
While the full details can be found in Refs.
\cite{Itamar,Itamar1,Itamar2}, here we just briefly explain
each of the terms. The first term of $f_{\rm{pol}}$ accounts for
chain elasticity, the second  describes the excluded volume
interaction between monomers, where $\it{v}$ is the second virial
coefficient. The third  accounts for the coupling with a
reservoir with bulk polymer concentration   $\phi_{\rm b}^2=c_{\rm b}$ and chemical
potential  $\mu$. The
$f_{\rm{ion}}$ contribution to the free energy takes into account
the entropy of small ions and their chemical potential $\mu^\pm$.
Lastly, $f_{\rm el}$ is the electrostatic free energy. Its first
term  is the interaction energy between the electrostatic
potential and the charged objects; namely, the small ions and
monomers. The last term is the self-energy of the electric field
 $-\frac{\varepsilon}{8\pi}\int \rm{d{\bf r}}\left|\nabla
\psi\right|^2$.

Minimizing the free energy with respect to $\psi$,  $\phi$,
$c^+$, $c^-$, and using the bulk boundary conditions: $\psi
\left(x\rightarrow \infty\right)=0$, $\phi \left(x\rightarrow
\infty\right) = \phi_{\rm b}$, $c_{\rm
b}^-=c_{\rm{salt}}+f\phi_{\rm b}^2$ and $c_{\rm
b}^+=c_{\rm{salt}}$, the profile equations of Ref. \cite{Itamar}
are reproduced:
\nolinebreak[2]
\begin{equation}
  c^-=\left(c_{\rm{salt}}+f\phi_{\rm b}^2\right)\e^{\beta e \psi}
 \label{cpluseq}
\end{equation}
\begin{equation}
  c^+=c_{\rm{salt}}\e^{-\beta e \psi}
 \label{cminuseq}
\end{equation}
\begin{equation}
  \nabla^2\psi= \frac{8\pi e c_{\rm{salt}}}{\varepsilon} \sinh {\beta e\psi} +
    \frac {4\pi e}{\varepsilon}\left(\phi_{\rm b}^2 f\e^{\beta e\psi}-
    f\phi^2\right)
 \label{PoissonBoltzmanmod}
\end{equation}
\begin{equation}
  \frac{a^2}{6}\nabla^2\phi={\it{v}}\left(\phi^3-\phi_{\rm b}^2\phi\right)+\beta
    f e\psi\phi ~,
 \label{Edwardsmod}
\end{equation}
where $\beta=1/k_{\rm B}T$ is the inverse of the thermal energy $k_B T$.
Equations~(\ref{cpluseq}) and
(\ref{cminuseq}) show that
the small ions obey Boltzmann statistics, while
Eq.~(\ref{PoissonBoltzmanmod}) is
the Poisson equation where the salt ions, counterions and monomers can
be regarded as the sources of
the electrostatic potential. Equation~(\ref{Edwardsmod}) is the mean-field
(Edwards)
equation for the
polymer order parameter $\phi({\bf r})$, taking into account the excluded volume
interaction and external electrostatic potential $\psi({\bf r})$.

The adsorption onto a flat, homogeneous and charged surface placed
at $x=0$ depends only on the distance $x$ from the surface. In
this case the above equations can be reduced to two coupled
ordinary differential equations. Defining  dimensionless
variables $\eta \equiv \phi/\phi_{\rm b}$ and $y\equiv \beta e \psi$, Eqs.
(\ref{PoissonBoltzmanmod}) and (\ref{Edwardsmod}) then read:
\nolinebreak[2]
\begin{equation}
  \frac{{\rm d}^2y}{{\rm d}x^2}=\kappa^2\sinh y +k_m^2 \left(\e^y-
    \eta^2 \right)\\
 \label{normPBmod}
\end{equation}
\begin{equation}
  \frac{a^2}{6}\frac{{\rm d}^2\eta}{{\rm d}x^2}=v\phi_{\rm b}^2
    \left(\eta^3-\eta \right)+f y \eta ~,
 \label{normEdwardsmod}
\end{equation}
where $\kappa^{-1}=\left(8\pi l_{\rm B} c_{\rm{salt}}\right)^{-1/2}$ is the
Debye-H\"{u}ckel screening length, determining the exponential decay of the potential
due to the added salt. Similarly,
$k_m^{-1}=\left(4\pi l_{\rm B} \phi_{\rm b}^2 f\right)^{-1/2}$ determines the
exponential decay due to the counterions. The Bjerrum length
is defined as $l_{\rm B}=e^2 / \varepsilon k_{\rm B} T$. For water
with dielectric constant $\varepsilon=80$, 
at room temperature, $l_{\rm B}$ is equal to about $7$\AA.
Note that the actual decay of the electrostatic potential is determined by a combination of
salt, counterions, and polymer screening effects.

The solution of Eqs. (\ref{normPBmod}) and (\ref{normEdwardsmod})
requires four boundary conditions. Two of them are the boundary values
in the bulk, $x\rightarrow \infty$: $\eta \left(x\rightarrow\infty\right)=1$ and
$ y\left(x\rightarrow\infty\right)=0$, while the other two are the boundary
conditions on the $x=0$ surface. In this article we use either constant surface charge density
(Neumann boundary conditions) or constant surface potential (Dirichlet boundary conditions).
For the former,
${\rm d}y/{\rm d}x\vert_{x=0}=-4\pi\sigma e/\varepsilon k_{\rm B} T=-4\pi(\sigma/e)l_{\rm B}$,
where $\sigma$ is the surface charge density.
For the latter, the surface potential is held fixed with a value:
$y\left(0\right)=y_s$.
The other boundary condition for the polymer concentration $\phi$
is taken as a non-adsorbing surface. Namely, $\phi\left(0\right)=0$.
Note that far from the surface, $x\rightarrow\infty$, Eqs.
(\ref{normPBmod}) and
(\ref{normEdwardsmod})
already satisfy
the boundary condition: $y=0$ and $\phi=\phi_{\rm b}$ (or $\eta=1$).

Equations (\ref{normPBmod}) and (\ref{normEdwardsmod}) are two
coupled non-linear differential equations that do not have a known
analytical solution. The numerical solutions of these equations
for low salt conditions were presented in Ref. \cite{Itamar} and
are reproduced here on Figs. 1 and 2, using a different numerical
scheme. The numerical results have been obtained using the
relaxation method \cite{NR}  based on a linearization procedure
done on a discrete one-dimensional grid. Then, the equations are transformed to a
set of algebraic equations for each grid point. The sum of the
absolute  difference between RHS and LHS over all grid points is
minimized iteratively until convergence of the numerical procedure
is achieved.

In calculating the numerical profiles of Figs.~1 and 2  we assume positively
charged polymers and a constant negative surface potential. In Fig.~1a
the reduced electrical
potential $y=\beta e \psi$ profile is shown as a function of the distance
from the $x=0$ surface. Similarly, in Fig.~2a the monomer
rescaled concentration profile $c(x)/c_{\rm b}$, is shown.
In both figures  a constant surface potential
boundary condition is imposed. The different curves correspond to different surface
potentials $y_s$, monomer charge fractions $f$ and monomer size
$a$. From the numerical profiles of the electrostatic potential
and monomer concentration it can be  clearly seen that there is a
distinct peak in both profiles. Although they do not occur 
exactly at the same distance from the surface, the corresponding 
peaks in Fig. 1 and 2
vary in a similar fashion
with system parameters. The peak in the concentration (Fig.~2) marks a PE
accumulation at the surface and is regarded as a signature of
adsorption.
The peak in the potential (Fig.~1) marks an
over-compensation of surface charges. At the peak of $\psi(x)$, the electric
field vanishes, $E=-{{\rm d}\psi}/{\rm d}x=0 $, meaning that the
integrated charge density from the surface up to this distance
exactly balances the surface charge. 


\section{Scaling Estimate of the Adsorption Layer: Counterion only Case}
\label{SCale}
\linespread{1.6}

So far numerical solutions within mean-field theory,
Eqs. (\ref{normPBmod}) and (\ref{normEdwardsmod}), have been described. We  proceed by
presenting simplified scaling
arguments,  which are in agreement with the numerical mean-field results.
Note that
the treatment here does not capture any correlation effect which goes
beyond mean-field.
The concept of polymer ``blobs" can be useful in order
to describe PE adsorption, where such polymer blob
can be regarded as a macro-ion adsorbing on a charge surface. The
blob size is determined by taking into account
the polymer connectivity and entropy
as well as the interaction with the charged surface.  A single
layer of adsorbing blobs is assumed instead of the full
continuous PE profile as obtained from the
mean-field equations. Therefore, the blob size characterizes the
adsorption layer thickness.

\subsection{Fixed Surface Charge Density}
\label{sigmaads}

The two largest contributions to the PE adsorption free energy
are the electrostatic attraction with the surface and the chain entropy
loss due to blob formation.  

For simplicity, the electrostatic attraction of the
monomers with the surface is assumed to be larger than the
monomer excluded volume and monomer-monomer electrostatic
repulsion.
With this assumption the chain has a Gaussian behavior inside
each surface blob, $D\sim a g^{1/2}$,  where $g$ is the number of
monomers in a blob of size $D$, as is shown schematically in
Fig.~3. The entropy loss of the chain balances the surface-monomer
attraction. As a result the blob attraction with the surface is
of order $k_B T$. It is now easy to get an estimate of the blob
size $D$:
\nolinebreak[2]
\begin{eqnarray}
   f g \frac{\scharge}{\varepsilon} e^2 D \simeq k_{\rm B} T
 \label{gsigmaeq} \\
   D\simeq ag^{1/2} \simeq \left(\frac{a^2}{l_{\rm B} f \scharge}\right)^{1/3}
 \label{Dscharge} \\
  g \simeq  \left(l_{\rm B} f a  \scharge\right)^{-2/3}~,
 \label{gscharge}
\end{eqnarray}
using a rescaled surface density $\scharge \equiv
\left|{\sigma}/{e}\right|$.
These results are in agreement with those describing the
statistics of single-chain adsorption \cite{Wiegel,Borisov}.

 The assumption that the electrostatic attraction to the surface is larger
than the monomer-monomer electrostatic repulsion and excluded
volume can now be checked self-consistently, yielding two
conditions: $ \scharge \gg  fa^{-2}$ and $ f \gg {\it
v}^3/(a^{10}\scharge l_B)$.

The average monomer concentration (per unit volume) in the
adsorption layer $c_m$, is the blob concentration in the
adsorption layer $n_0$, times the number of monomers per blob
$g$, yielding
 $c_m=n_0 g$.
It is now possible to get an estimate of the blob concentration
per unit volume in the adsorption layer, $n_0$, by assuming that
the adsorbed layer neutralizes the surface charges up to a
numerical prefactor of order unity. Hence, $n_0\simeq
\scharge/Dfg$. This assumption, which is in agreement with our numerical
solutions, leads to
\nolinebreak[2]
\begin{eqnarray}
  n_0& \simeq& l_{\rm B}\scharge^2
 \label{n0sig} \\
  c_m &=&n_0 g \simeq (l_{\rm B} a^{-2}f^{-2}\scharge^{4})^{1/3} \sim \scharge^{4/3}f^{-2/3} ~.
 \label{cmscharge}
\end{eqnarray}
Equation (\ref{n0sig}) is just the Graham equation \cite{Israelachvili}
relating the surface charge density
with the counterion density at the surface vicinity. The only difference is that
the counterions
are replaced by the charged polymer blobs.  Furthermore, Eq. (\ref{cmscharge})
is in accord with the results of Ref \cite{review}.

The total amount of PEs in the adsorption layer is
$  \Gamma  \simeq c_m D  ={\scharge}/{f}$. In other words,
the overall  polymer charge in the  adsorption layer (up to a numerical prefactor) is
$  f\Gamma \simeq \scharge$.
This is just another way to phrase the charge neutralization
by the PEs mentioned above.

\subsection{Fixed Surface Potential}
\label{ysads}

Using the boundary condition of a fixed surface potential
$\psi=\psi_s$, the scaling laws for $D$ and $g$ can be obtained in
a similar fashion as was done in Sec. \ref{sigmaads}. Alternatively,
one can (in the absence of salt) relate the surface potential to the 
surface charge density by $\psi_s\simeq \sigma D/\varepsilon$.

The adsorption energy of a blob of charge $gfe$ onto a surface
held at potential $\psi_s$ is just $gfe\psi_s$. Requiring that
this energy is of order of $k_BT$ we obtain in analogy to Eqs.
(\ref{gsigmaeq})-(\ref{gscharge}):
\nolinebreak[2]
\begin{eqnarray}
  gfe\left|\psi_s\right| \simeq k_{\rm B} T \label{gpsieq} \\
  D \simeq ag^{\half}\simeq \frac{a}{\sqrt{f \ys}} \label{Dys}\\
  g \simeq \frac{1}{f\ys } \label{gys}
\end{eqnarray}
Together with the neutralization condition $n_0\simeq \scharge/(D f g)$ it yields:
\begin{eqnarray}
  n_0 \simeq  \frac{f\ys^3}{l_{\rm B} a^2}
 \label{n02} \\
  c_m= n_0 g \simeq \frac{\ys^2}{l_{\rm B} a^2}.
 \label{cmys}
\end{eqnarray}
Note that the above results are in accord with the
ones previously derived in Ref. \cite{Itamar}.

Just like in Sec. III.A  the self-consistent check can be repeated here
for the dominance of the surface-monomer
interactions, yielding $\ys \gg   f^{1/3}l_B^{2/3}a^{-2/3}$ and $f \gg {\it{v}^2}/{a^6 \ys}$.
This condition  has been verified, in addition, by examining numerically the mean-field
adsorbing profiles.

The overall charge of the polymer in the adsorbed layer is then:
\nolinebreak[2]
\begin{eqnarray}
  \Gamma  \simeq c_m D \simeq
    \ys^{3/2}f^{-1/2} l_{\rm B}^{-1}a^{-1} \sim \ys^{3/2}f^{-1/2}
 \label{gammays}\\
f\Gamma \simeq \ys^{3/2}f^{1/2} l_{\rm B}^{-1} a^{-1}\simeq
         \frac{\ys}{l_{\rm B} D} \simeq \left|\frac{d\psi}{dx}\right|_{x=0} \simeq \scharge
 \label{delsigsig}
\end{eqnarray}
which again verifies that the adsorbed amount scales like the surface charge.

The numerical results of the mean-field equations for constant surface
potential $y_s$ condition and in the low
salt regime ($c_{\rm salt}=0.1\,$mM)  are consistent with this scaling picture, as can be
seen in Figs.~1 and 2. In Fig.~1b  the rescaled potential $y/|y_s|$ is
plotted
in terms of a rescaled distance: $x/D$, with $D$ taken from Eq. (\ref{Dys}).
In Fig.~2b the concentration profile is rescaled by $c_m$, Eq.~(\ref{cmys}),
and plotted in terms of the same rescaled distance $x/D$. The figures show clearly
data collapse of the two profiles, indicating that the characteristic adsorption
length $D$ is indeed given by the scaling predictions. Note that the agreement with
the scaling argument occurs as long as the system stays in the low salt limit.
The other limit of high salt is discussed next.

\section{The Adsorption -- Depletion Transition in presence of Added Salt}
\label{Phaset}
\linespread{1.6}

The same numerical procedure outlined in Sec. II, is used
to find when the chains stop adsorbing and instead will
deplete from the surface. This is not a sharp transition but rather
a crossover which is seen by calculating numerically the PE surface excess, as
depicted in Fig.~4.
The profiles  were obtained by solving numerically the
differential equations for several values of $f$ near the
adsorption--depletion transition using a fixed surface potential
boundary condition. For salt concentration of about
$c_{\rm salt}^* \simeq 0.16 \ys f/(l_{\rm B}a^2)$ (solid line in Fig.~4),  the
figure show the disappearance of the
concentration peak. Namely, a depletion--adsorption crossover.

The dependence of $\Gamma=\int_0^\infty {\rm d}x \left(\phi^2-\phi_b^2\right)$
on $c_{\rm salt}$ and $f$ for constant surface potential is presented in Fig.~5.
The place where $\Gamma=0$ indicates an adsorption--depletion transition, separating
positive $\Gamma$ in the adsorption regime from negative ones in the depletion regime. In Fig.~5a the
dependence of $\Gamma$ on $f$ is shown for several salt concentrations ranging
from low-  to high-salt conditions. For low enough $f$, $\Gamma<0$ indicates depletion.
As $f$ increases, a crossover to the adsorption region, $\Gamma>0$, is seen.
In the adsorption region, a peak in $\Gamma(f)$ signals the maximum adsorption amount at constant $c_{\rm salt}$.
As $f$ increases further, beyond the peak, $\Gamma$ decreases as $1/\sqrt{f}$.
Looking at the variation of $\Gamma$ with salt, as $c_{\rm salt}$ increases, the
peak in $\Gamma(f)$ decreases and shifts to higher values of $f$. For very large amount of salt, {\it e.g.,}
$c_{\rm salt}=0.5$\,M,
the peak occurs in  the limit $f\to 1$.
In Fig. 5b, we plot $\Gamma(c_{\rm salt})$ for several $f$ values. The adsorption regime
crosses over to depletion quite sharply as $c_{\rm salt}$ increases,
signaling the adsorption-depletion transition. The salt concentration at
the transition, $c_{\rm salt}^*$, increases with the charge fraction $f$.
The dependence of $\Gamma$ on $c_{\rm salt}$ and $f$ for constant surface
charge density is plotted in Fig.~6. Both salt 
and $f$ dependences show a similar behaviour 
to those shown in Fig.~5 for constant surface potential.

The numerical phase diagrams supporting the
adsorption--depletion transition are presented in Fig.~7
for constant surface charge conditions.
The phase diagrams were obtained by solving numerically the
mean-field equations. We
scanned the $(f, c_{\rm salt})$ parameter plane for 50 values of
$f$ between $0.01<f<1$ (Fig.~7a) and the $(\scharge, c_{\rm salt})$
plane for 50 values of $\scharge=|\sigma/e|$ between
$10^{-5}$\,[\AA$^{-2}$] $< \scharge < 10^{-4}$\,[\AA$^{-2}$]
(Fig.~7b).  From
the log-log plots it can be seen that  the adsorption--depletion  transition  is described
extremely well by a  line of slope $2/3$ in both Fig. 7a and 7b.
Namely, at the transition $c_{\rm salt}^*\sim f^{2/3}$
for fixed $\scharge$ and $c_{\rm salt}^*\sim \scharge^{2/3}$ for fixed $f$.

To complete the picture, the
adsorption--depletion transition is also presented in Fig.~8 for constant surface potential.
The phase diagrams are obtained by solving numerically the
differential equations. We
scanned the $(f, c_{\rm salt})$ parameter plane for 50 values of
$f$ between $0.01<f<1$ (Fig.~8a) and the $(\ys, c_{\rm salt})$
plane for 50 values of $\ys$ between $0.1<\ys<1.0$ (Fig.~8b).  From
the figure it is apparent that  the adsorption--depletion  transition line fits
quite well a line of slope 1.0 in both Fig. 8a and 8b plotted on a log-log
scale. Namely, $c_{\rm salt}^*\sim f$
for fixed $y_{\rm s}$, and $c_{\rm salt}^*\sim y_{\rm s}$ for fixed $f$.

These scaling forms of  $c_{\rm salt}^*$ at the transition can be
explained using the simplified scaling arguments introduced in Sec. III.

\subsection{Scaling for Fixed Surface Charge}
\label{sigdepsubsec}

If  the blobs are taken as charged spheres, the mere existence of an adsorption
process requires that the attraction of the monomers to the surface
persists for all charges  up to distances $D$ from the charged surface.
For high ionic strength solutions, the electrostatic potential at distance $x$
for a charged surface can be approximated by the linearized Debye-H\"uckel potential:
\nolinebreak[2]
\begin{equation}
  y\left(x\right)= 4\pi \scharge l_{\rm B} \kappa^{-1} \e^{-\kappa x}~.
 \label{schargekappa}
\end{equation}
This is valid as long as the potential
is low enough, $y\le 1$. The adsorption picture requires that the exponential
decay of the potential will not vary substantially inside a region of size $D$
comparable to the size of surface blobs, $y(D)\simeq y_s$.
Then,  the exponential decay in Eq.~(\ref{schargekappa}) yields
\nolinebreak[2]
\begin{equation}
   \kappa D < 1 ~.
 \label{depletion1}
\end{equation}
Namely, the Debye-H\"uckel screening length is smaller than the
adsorption layer thickness, $D$.
Using Eq.~(\ref{Dscharge}) this yields:
\nolinebreak[2]
\begin{equation}
  c_{\rm{salt}} < \scharge^{2/3} f^{2/3}l_{\rm B}^{-1/3}a^{-4/3}
 \label{depletion_condition_scharge}
\end{equation}
The crossover between adsorption and
depletion will occur when
$c_{\rm{salt}}^* \simeq (\scharge^{2} f^{2}l_{\rm B}^{-1}a^{-4})^{1/3} $, in accord with
Refs. \cite{Wiegel,Muthukumar,Dobrynin}, and with the numerical results
discussed above and presented in Fig.~7.

\subsection{Scaling for Fixed Surface Potential}
\label{ysdepsubsec}

For the boundary condition, $\psi=\psi_s$, the potential decay
from the surface can be approximated to be:
\nolinebreak[2]
\begin{equation}
  y\left(x\right) = y_s \e^{-\kappa x} \\
 \label{yskappa}
\end{equation}
and the same consideration as in Eq. (\ref{depletion1}) and
(\ref{depletion_condition_scharge}) gives:
\nolinebreak[2]
\begin{equation}
c_{\rm{salt}} < \frac{\ys f}{l_{\rm B} a^2}.
 \label{depletion_condition_ys}
\end{equation}
Namely, we expect an adsorption--depletion transition to occur
for $c_{\rm salt}^* \simeq \ys f/(l_{\rm B}a^2)$, in the case
of a fixed surface potential. This supports the numerical results
as presented in Fig.~8.

\section{Discussion}
\label{comparison}
\linespread{1.6}

   We have presented numerical calculations of the
mean-field equations describing the
adsorption of PE chains onto charged surfaces, including multi-chain
interactions. The main finding is
the existence of an adsorption--depletion transition
in presence of added salt or weakly charged chains.
The numerical results are discussed in terms of
simple scaling arguments describing the adsorption of
PEs. The salt concentration at the
adsorption--depletion transition scales like
$c_{\rm salt}^* \sim f\ys$ for fixed surface potential and
$c_{\rm salt}^* \sim \left(f\scharge\right)^{2/3}$ for  fixed surface charge density.
Within the scaling picture,
the condition for depletion is the same as for a single chain, in agreement with our mean-field solutions.

We briefly summarize the main approximations
of our mean-field and scaling results.
A non-adsorbing surface is used as the polymer boundary
condition. However, if the surface has a strong
non-electrostatic affinity for the PE chains, the electrostatic
contribution does not have to be the dominant one. The
method also assumes Gaussian blobs within
mean-field theory. In a more refined theory,
excluded volume interactions as well as
lateral correlation in the blob-blob interactions will alter the adsorption behavior.
When the surface charge (or potential) is high enough, the
blob size $D$ can become comparable with the monomer size $a$, and
the PE chains will lay flat on the surface. Further investigations might
be necessary to address in more detail the above points.
It will  also be interesting to extend our results
to geometries other than the planar charged surface.

Several authors have addressed the problem of adsorption onto
surfaces either of a single chain \cite{Borisov} or multiple
chains \cite{Dobrynin} using similar arguments of blobs.
In another approach, a Flory-like free energy \cite{Itamar} was
introduced using the assumption of a single characteristic length
scale. The latter gave adsorption-layer scaling laws as in
Eqs.~(\ref{Dys}) and (\ref{cmys}), but did not find the depletion
criterion. Instead, an adsorption length scale and a
characteristic concentration were predicted for the high-salt
regime. We show here, using both   numerical
calculations and scaling arguments, that the high-salt regime
does not exist because it
is preempted by a PE depletion.

\vskip 2truecm

{\em Acknowledgments.~~~~} It is a pleasure to thank
G. Ariel, I. Borukhov, Y. Burak, E. Katzav and H. Orland for useful discussions
and comments. DA acknowledges support from the Israel Science
Foundation under grant No. {210/02} and the Alexander von Humboldt 
Foundation. RRN acknowledges support from the 
Deutsche Forschungsgemeinschaft (DFG, German-French Network) 
and the Fonds der Chemischen Industrie.


\newpage
\section*{Figure Captions}

\begin{itemize}

 \item[{\bf Fig. 1}]
    (a) Numerical profiles of the rescaled electrostatic  potential
    $y=\beta e \psi$ as function of the distance from the surface $x$
     using Eqs.~(\ref{normPBmod})  and (\ref{normEdwardsmod}), and constant
     surface potential.
     The solid line
        is for $a=5$\AA, $f=1, y_s=-1.0$, the dotted line for
        $a=5$\AA, $f=1, y_s=-0.5$, the dashed line is
        for $a=10$\AA, $f=1, y_s=-0.5$,  and the
         dash-dot line for $a=5$\AA, $f=0.1, y_s=-0.5$.
        All profiles have $c_{\rm salt}=0.1$\,{mM}, $\phi_{\rm b}^2=10^{-6}$
        \AA$^{-3}$, $v=50$\AA$^3$, $\varepsilon =80,\, T=300$\,K. The profiles
     reproduce those of Ref. \cite{Itamar} using a different numerical scheme. (b)
     Same profiles as in part (a) but
    in rescaled variables: $x/D$ and $y/\ys$.

\item[{\bf Fig. 2}]
    (a) The concentration profile $c(x)/c_b=\phi^2(x)/\phi_{\rm b}^2$ for the numerical
     calculations specified
        in Fig.~1.
        The profiles reproduce those of Ref. \cite{Itamar}. (b) Same
     as in part (a)
        but in rescaled variables: $x/D$ and $c(x)/c_m$.

\item[{\bf Fig. 3}]
    A schematic drawing of polyelectrolyte adsorption onto flat surfaces and
    formation of Gaussian surface blobs each of size $D$ and having $g$ monomers.
    The monomer size is $a$.

\item[{\bf Fig. 4}]
    Numerical polyelectrolyte concentration profiles exhibiting the transition from 
    adsorption to depletion. The dashed line corresponds to
    $f=0.12$, the dot--dash line to $f=0.1$, the solid line to $f=0.09$,
    and the dotted line to $f=0.08$. All profiles have
    $\ys=0.5$, $\phi_b^2=10^{-6}$\,\AA$^{-3}$, $v=50$\,\AA$^3$, $a=5$\,\AA,
    $c_{\rm salt}=70\,$mM.  The adsorption--depletion transition is found
    to occur for $f=0.09$, corresponding
    to $c_{\rm salt}^*\simeq 0.16{\ys f}/{l_B a^2}$.

 \item[{\bf Fig. 5}] (a) Surface excess of PE adsorption, $\Gamma$, as
function of the chain charged fraction $f$, for several salt
concentrations: 1.0\,mM (solid line), 10\,mM (dashed line), 0.1\,M
(dash-dot line), 0.5\,M (dots), and for constant surface potential. 
As the salt concentration increases, the
peak in $\Gamma$ shifts to higher $f$ values and disappears for $c_{\rm
salt}=0.5$\,M. The depletion-adsorption transition occurs for $\Gamma=0$.
(b) Surface excess as function of salt concentration, $c_{\rm salt}$, for
several $f$ values: f=0.03 (dots), 0.1 (dashes), 0.3 (dot-dash), 1.0
(solid line). $\Gamma$ is almost independent of $c_{\rm salt}$ for low
salt concentrations in the adsorption region. It is then
 followed by a steep descent into a depletion region at a threshold value.
 Other parameters used are: $y_s=-1.0$, ${\it v}=50$\AA$^3$, 
 $\phi_b^2=10^{-6}$\AA$^{-3}$, $a=5$\AA, $T=300$K and $\varepsilon=80$. 
 
 \item[{\bf Fig. 6}] (a) Surface excess of PE adsorption, $\Gamma$, as
function of the chain charged fraction $f$, for several salt
concentrations: 4.0\,mM (solid line), 8.0\,mM (dashed line), 21\,mM
(dash-dot line), 63\,mM (dots), and for constant surface charge density. 
As the salt concentration increases, the
peak in $\Gamma$ shifts to higher $f$ values and disappears for $c_{\rm
salt}=63$\,mM. The depletion-adsorption transition occurs for $\Gamma=0$.
(b) Surface excess as function of salt concentration, $c_{\rm salt}$, for
several $f$ values: f=0.1 (dots), 0.2 (dashes), 0.45 (dot-dash), 1.0
(solid line). $\Gamma$ is almost independent of $c_{\rm salt}$ for low
salt concentrations in the adsorption region. It is then
 followed by a steep descent into a depletion region at a threshold value.
 Other parameters used are: $\sigma/e=-10^{-4}$\AA$^{-2}$, ${\it v}=50$\AA$^3$, 
 $\phi_b^2=10^{-6}$\AA$^{-3}$, $a=5$\AA, $T=300$K and $\varepsilon=80$.

\item[{\bf Fig. 7}]
     Numerically calculated adsorption--depletion crossover diagram for constant surface
         charge condition.  In (a) the
         $(f,c_{\rm salt})$  parameter plane on  a log-log scale while $\scharge=|\sigma/e|$
         is held constant at $\scharge=10^{-3}$\,\AA$^{-2}$.
     The full squares represent the lowest salt concentration for which depletion is detected.
    The  least-mean-square fit to the data points gives a straight line with
    slope of $0.69\pm 0.02$.  The figure
          shows that the numerical results agree with a $2/3$ power
          law as predicted in Sec. IV.A, $c_{\rm salt}^* \sim f^{2/3}$. In (b)
      the crossover diagram in calculated numerically in the
      $(|\sigma/e|,c_{\rm salt})$ parameter plane on
         a log-log scale, while $f$ is fixed to be $f=0.1$
    The least-mean-square line has a slope of $0.71\pm0.02$,
          showing that the numerical results agree with a $2/3$ power
          law as predicted in Sec. IV.A, $c_{\rm salt}^* \sim \sigma^{2/3}$.

\item[{\bf Fig. 8}]
     Numerically calculated crossover diagram on
         a log-log scale for constant surface potential conditions. Notations and symbols are
         the same as in Fig.~6. In  (a) the
         $(f,c_{\rm salt})$ parameter plane is presented for constant $y_{\rm s}=-1.0$.
      The  least-mean-square fit
        has
         a slope of $1.00 \pm 0.02$, in excellent agreement with  the scaling arguments,
     $c_{\rm salt}^*\sim f$. In (b)
         the $(\ys,c_{\rm salt})$ parameter plane is presented, for constant $f=0.1$.
          The least-mean-square fit  has a slope of $1.04 \pm 0.02$,
     in agreement with
     scaling arguments, $c_{\rm salt}^*\sim \ys$.

\end{itemize}

\begin{widetext}

\newpage

\begin{figure}[tbh]
\epsfxsize=1.0\linewidth
\centerline{\hbox{ \epsffile{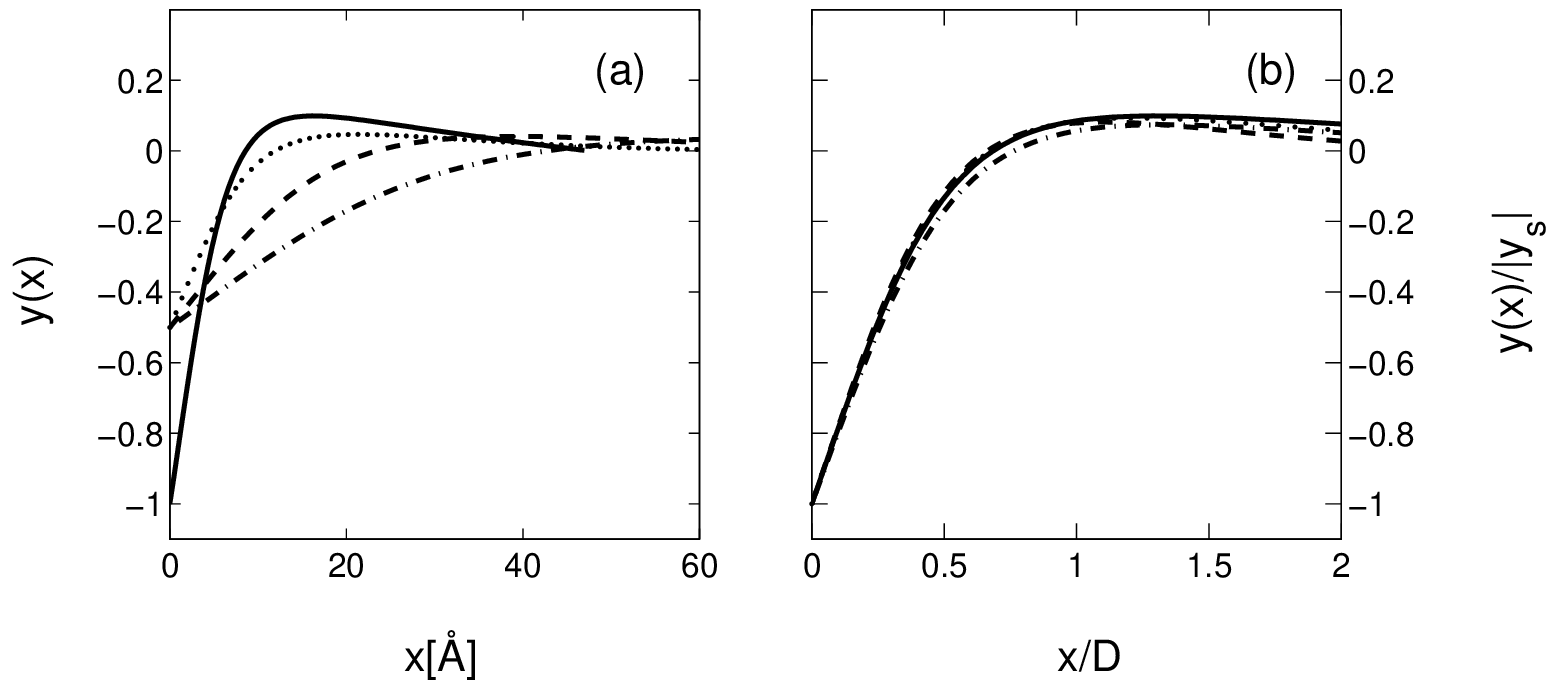} } }
\end{figure}
\centerline{\large Fig.~1, Shafir et al}
~~~~~
\bigskip
~
\bigskip
~
\bigskip

\begin{figure}[tbh]
\epsfxsize=1.0\linewidth
\centerline{\hbox{ \epsffile{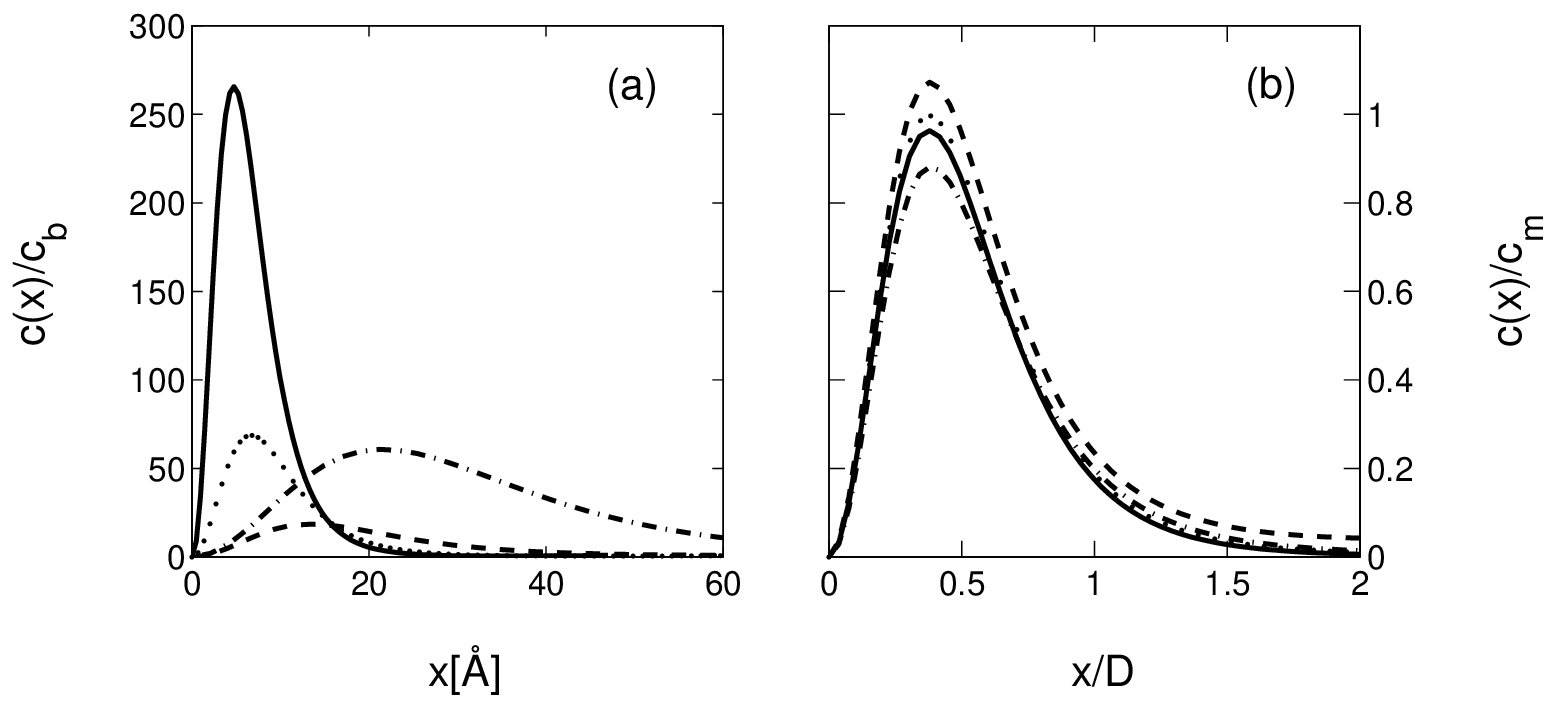} } }
\end{figure}
\centerline{\large Fig.~2, Shafir et al}

\newpage
\end{widetext}

\begin{figure}[tbh]
\epsfxsize=1.0\linewidth
\centerline{\hbox{ \epsffile{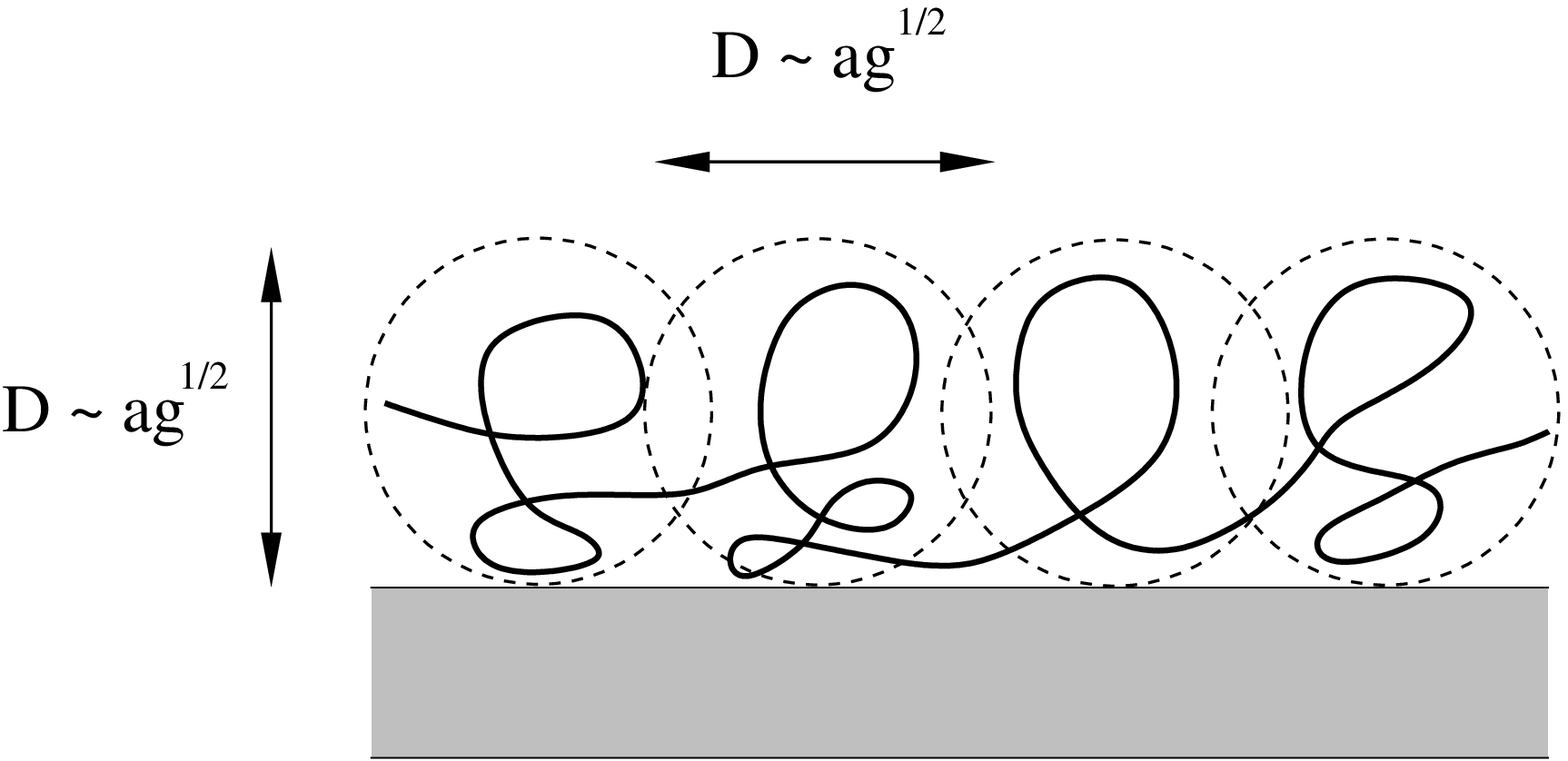} } }
\end{figure}
\centerline{\large Fig.~3, Shafir et al}

\newpage

\begin{figure}[tbh]
\epsfxsize=1.0\linewidth
\centerline{\hbox{ \epsffile{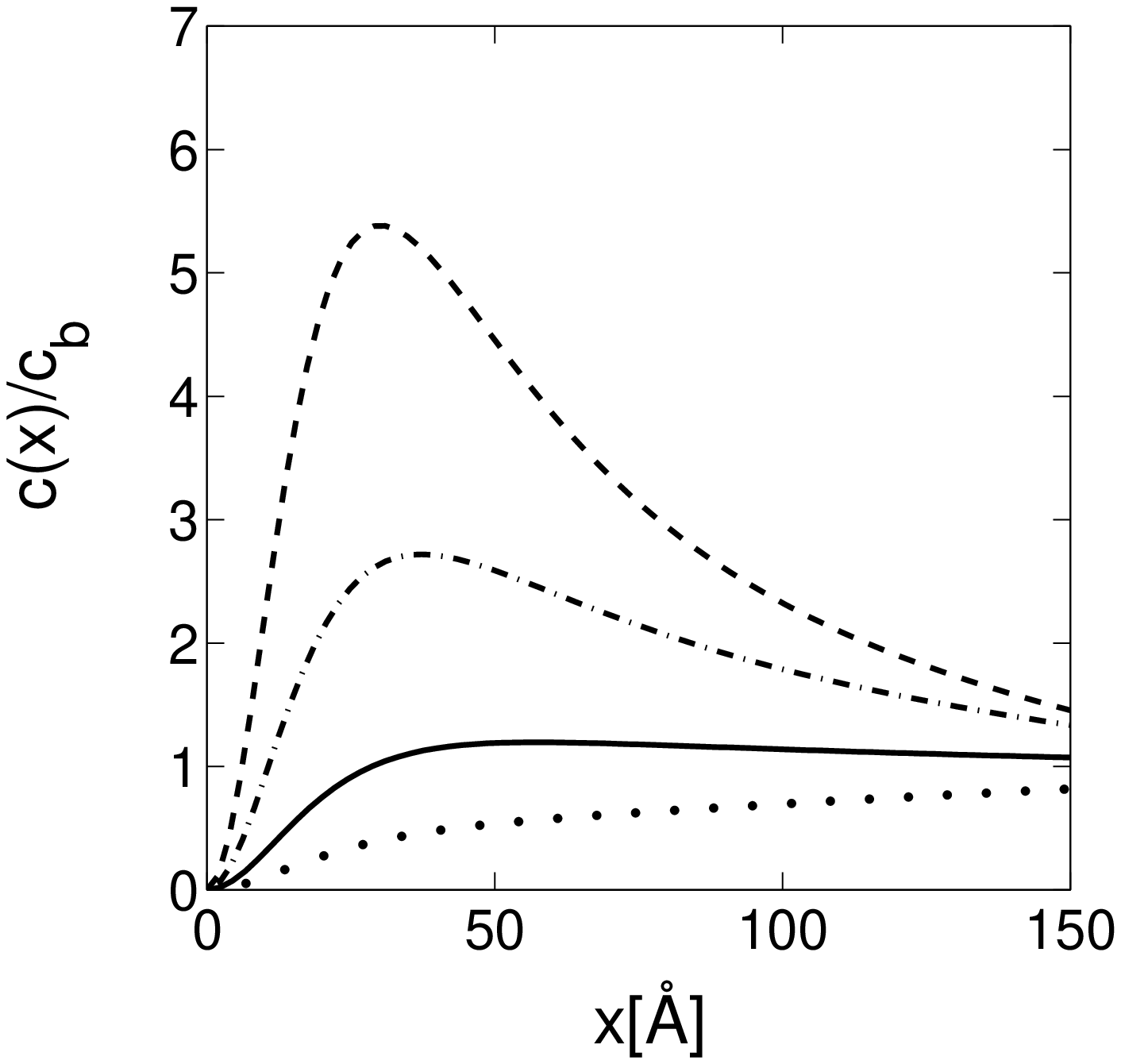} } }
\end{figure}
\centerline{\large Fig.~4, Shafir et al}

\newpage

\begin{widetext}

\begin{figure}[tbh]
\epsfxsize=1.0\linewidth
\centerline{\hbox{ \epsffile{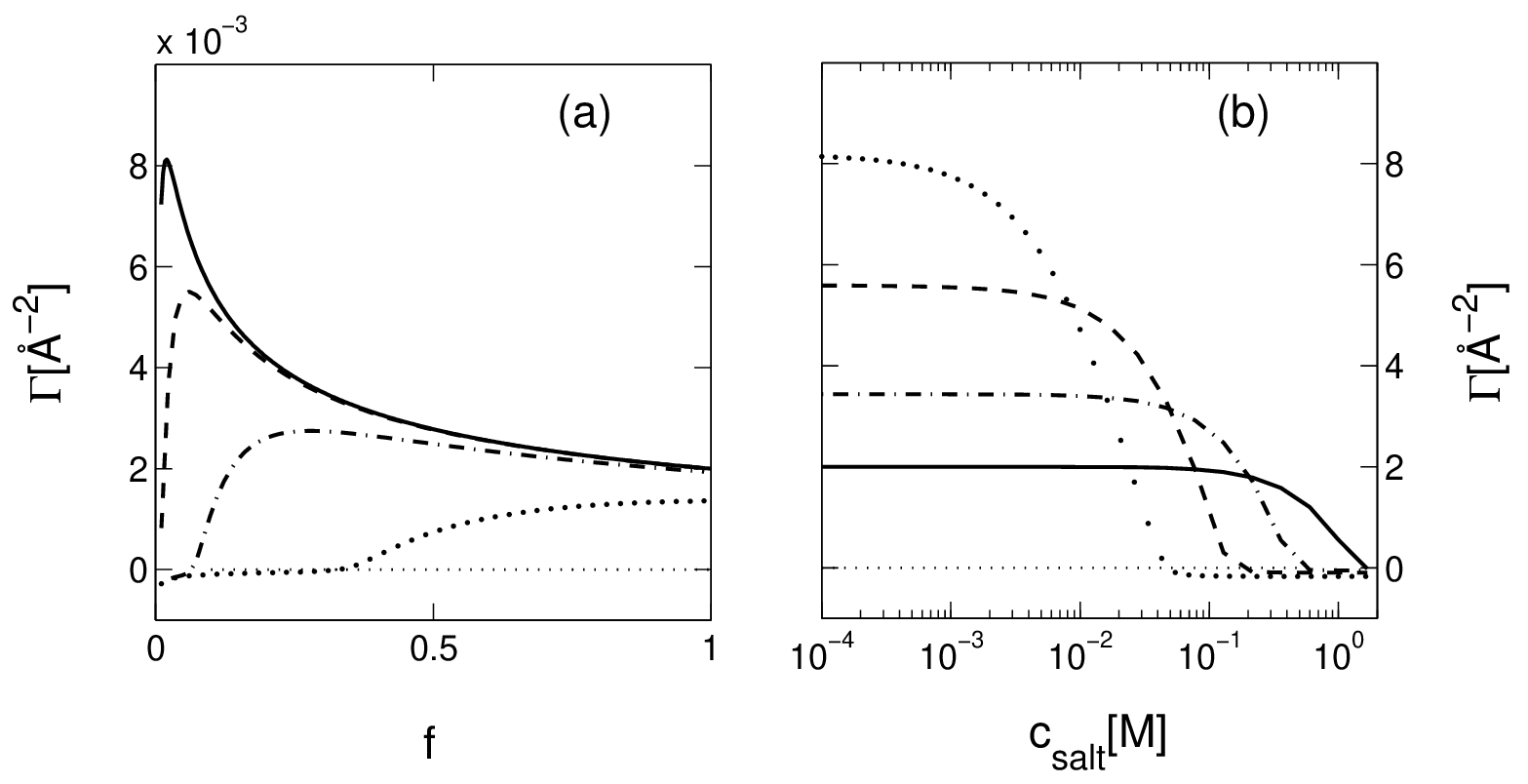} } }
\end{figure}
\centerline{\large Fig.~5, Shafir et al}
~~~~~
\bigskip
~
\bigskip
~
\bigskip

\begin{figure}[tbh]
\epsfxsize=1.0\linewidth
\centerline{\hbox{ \epsffile{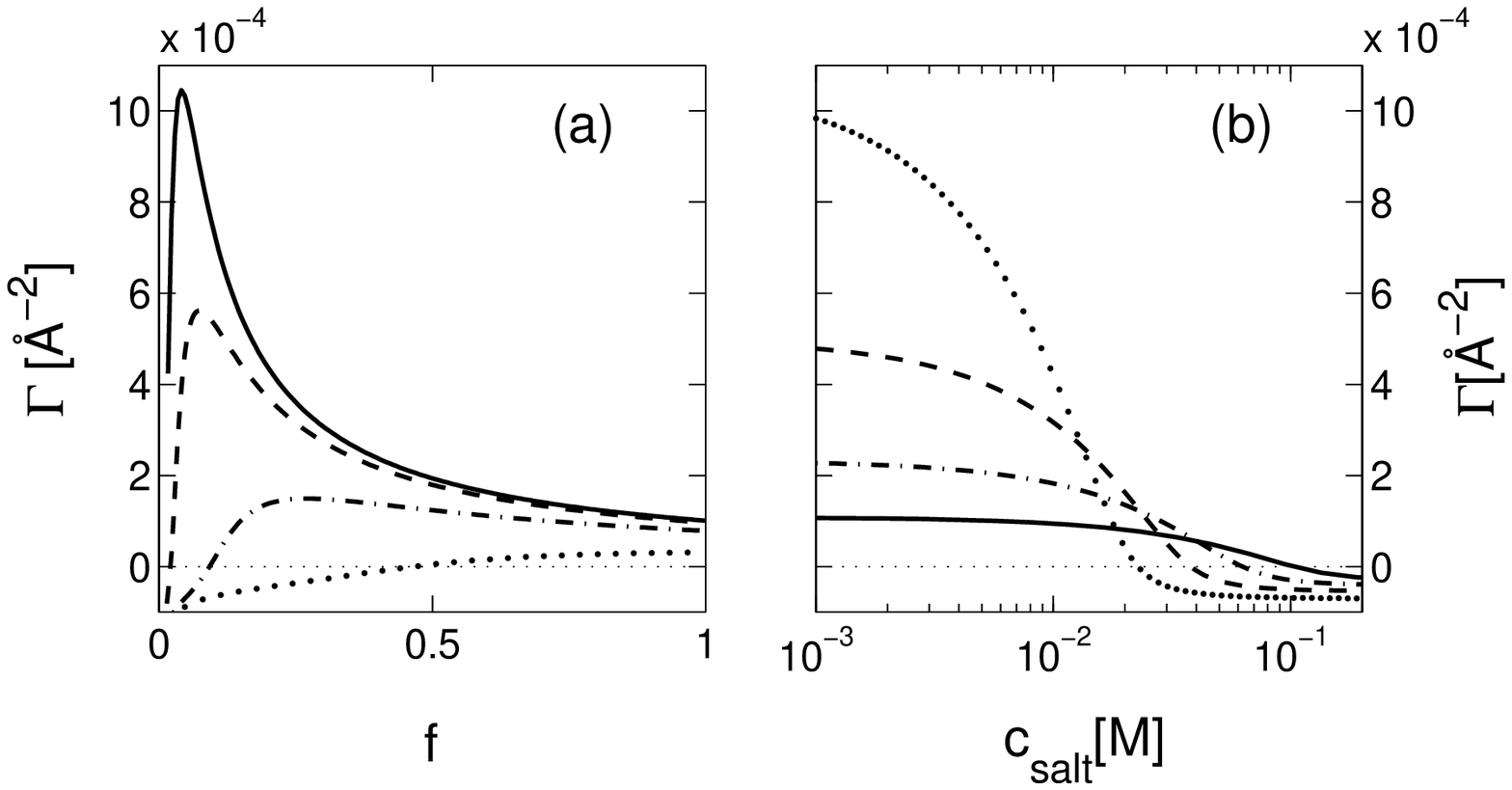} } }
\end{figure}
\centerline{\large Fig.~6, Shafir et al}

\newpage
\begin{figure}[tbh]
\epsfxsize=1.0\linewidth
\centerline{\hbox{ \epsffile{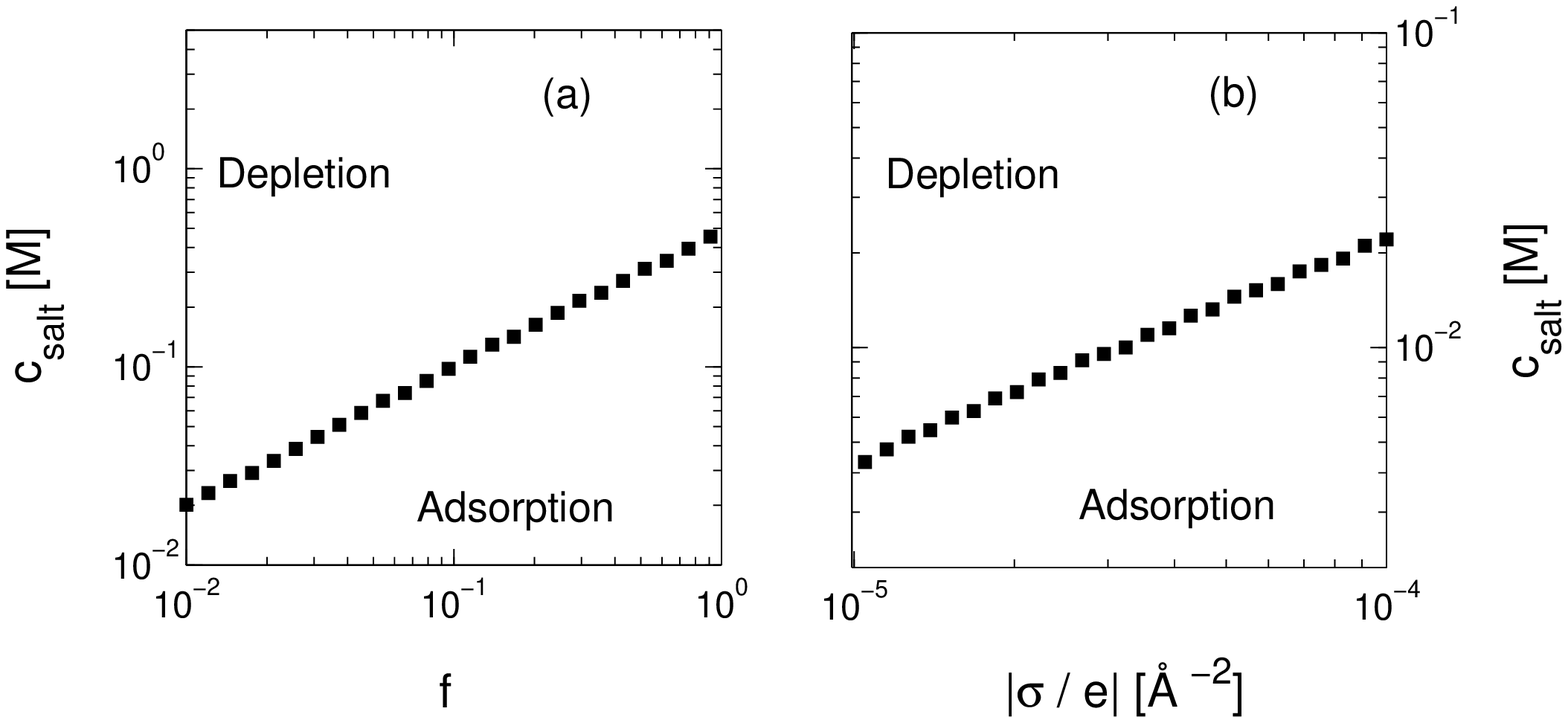} } }
\end{figure}
\centerline{\large Fig.~7, Shafir et al}

~~~~~
\bigskip
~
\bigskip
~
\bigskip

\begin{figure}[tbh]
\epsfxsize=1.0\linewidth
\centerline{\hbox{ \epsffile{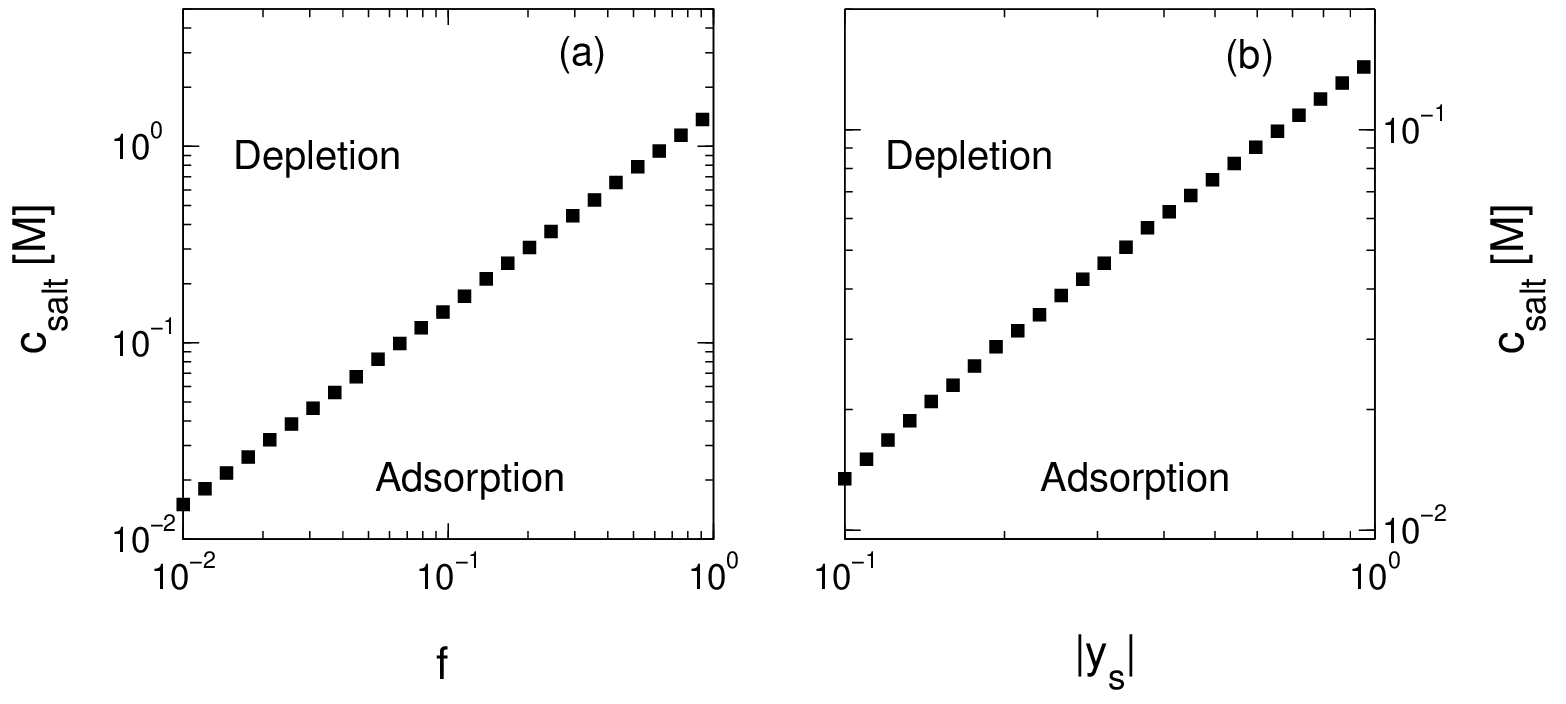} } }
\end{figure}
\centerline{\large Fig.~8, Shafir et al}
\newpage

\end{widetext}

\end{document}